\documentclass[acmsmall]{acmart}

\renewcommand\footnotetextcopyrightpermission[1]{}
\AtBeginDocument{%
  }

\begin{document}

\title{A Fine-Grained and Efficient Reliability Analysis Framework for Noisy Quantum Circuits}

\author{Jindi Wu}
\email{jwu115@depaul.edu}
\orcid{0000-0002-0489-0616}
\affiliation{%
  \institution{DePaul University}
  % \city{~Chicago}
  % \state{~Illinois}
  \country{~USA}
}

\author{Tianjie Hu}
\email{thu04@wm.edu}
\affiliation{%
  \institution{~College of William and Mary}
  % \city{~Williamsburg}
  % \state{~Virginia}
  \country{~USA}
}

\author{Qun Li}
\email{liqun@cs.wm.edu}
\affiliation{%
  \institution{~College of William and Mary}
  % \city{~Williamsburg}
  % \state{~Virginia}
  \country{~USA}
}

% \renewcommand{\shortauthors}{Wu et al.}

\renewcommand{\shortauthors}{Jindi Wu, Tianjie Hu, and Qun Li}

\begin{abstract}
Evaluating the reliability of noisy quantum circuits is essential for implementing quantum algorithms on noisy quantum devices. However, current quantum hardware exhibits diverse noise mechanisms whose compounded effects make accurate and efficient reliability evaluation challenging. While state fidelity is the most faithful indicator of circuit reliability, it is experimentally and computationally prohibitive to obtain. Alternative metrics, although easier to compute, often fail to accurately reflect circuit reliability, lack universality across circuit types, or offer limited interpretability. To address these challenges, we propose a fine-grained, scalable, and interpretable framework for efficient and accurate reliability evaluation of noisy quantum circuits. Our approach performs a state-independent analysis to model how circuit reliability progressively degrades during execution. We introduce the \textit{Noise Proxy Circuit (NPC)}, which removes all logical operations while preserving the complete sequence of noise channels, thereby providing an abstraction of cumulative noise effects. Based on the NPC, we define \textit{Proxy Fidelity}, a reliability metric that quantifies both qubit-level and circuit-level reliability. We further develop an analytical algorithm to estimate Proxy Fidelity under depolarizing, thermal relaxation, and readout error channels. The proposed framework achieves fidelity-level reliability estimation while remaining execution-free, scalable, and interpretable. Experimental results show that our method accurately estimates circuit fidelity, with an average absolute difference (AAD) ranging from 0.031 to 0.069 across diverse circuits and devices. 

\end{abstract}

%   <concept_desc>Theory of computation, Quantum information theory</concept_desc>



% \received{October 2025}
% \received[revised]{December 2025}
% \received[accepted]{January 2026}

\maketitle

\section{Introduction} \label{sec:intro}

Quantum computing has attracted significant attention in recent years due to its potential to solve certain problems that are intractable for classical computers \cite{nielsen2001quantum}. Various physical platforms have been proposed for building quantum computers, including superconducting qubits \cite{gambetta2017building}, trapped ions \cite{haffner2008quantum}, photonic systems \cite{abughanem2024photonic}, and neutral atoms \cite{henriet2020quantum}.
However, one of the most critical challenges in contemporary quantum computing is quantum noise \cite{clerk2010introduction}, which arises from the inherent instability of quantum systems and the imperfect control of quantum operations. Such noise significantly degrades computational reliability and limits the scalability of quantum processors. As a result, current quantum computers are commonly referred to as noisy intermediate-scale quantum (NISQ) devices \cite{bharti2022noisy}.
Although quantum error correction (QEC) \cite{lidar2013quantum} has been proposed to enable fault-tolerant quantum computing, its practical deployment remains infeasible on today’s NISQ hardware due to limited qubit resources and substantial operational overhead. Consequently, recent research efforts have largely focused on software-level techniques that aim to mitigate or reduce the impact of noise during quantum computation \cite{endo2018practical, strikis2021learning, endo2021hybrid, kim2023evidence, temme2017error}.

To minimize the impact of noise during circuit execution, noise-aware quantum circuit implementation, also referred to as circuit compilation optimization, has become an essential component of practical quantum computing workflows \cite{fosel2021quantum, ge2024quantum}. This process preserves the logical functionality of a circuit while adapting its structure to the characteristics of the target hardware, such as gate error rates, qubit connectivity, and operation durations, thereby improving execution reliability. After circuit execution, quantum error mitigation (QEM) techniques are often applied to recover ideal outputs from noisy measurement results using statistical or heuristic methods \cite{strikis2021learning, cai2023quantum, endo2018practical}. Beyond circuits that directly produce computational outputs, noise-aware implementation is also critical in other scenarios, including quantum state preparation without measurement and subcircuit execution in distributed quantum computing \cite{elliott2002building, felinto2006conditional, gisin2007quantum, cacciapuoti2019quantum}. Ensuring the reliability of such circuits is therefore fundamental to achieving accurate and scalable quantum computation.

Reliability analysis and evaluation play a central role in quantum circuit compilation optimization and related applications, as they guide the selection of circuit implementations that are more likely to execute reliably on a target quantum processor \cite{nishio2020extracting, wu2024detecting, wu2024q, ghaderibaneh2022efficient, 11303360}. 
However, evaluating the reliability of noisy quantum circuits remains highly challenging due to the diversity of quantum noise sources and the complexity of their interactions.
In a quantum circuit, errors may arise from imperfect gate operations, state preparation and measurement (SPAM) errors, amplitude damping, and dephasing. These noise mechanisms exhibit distinct characteristics and may interact nontrivially during circuit execution. As a result, their accumulated effects are difficult to model accurately, hindering precise assessment of how circuit reliability degrades under realistic hardware noise.

\begin{figure*}[t]
\centering
\includegraphics[width=1\linewidth]{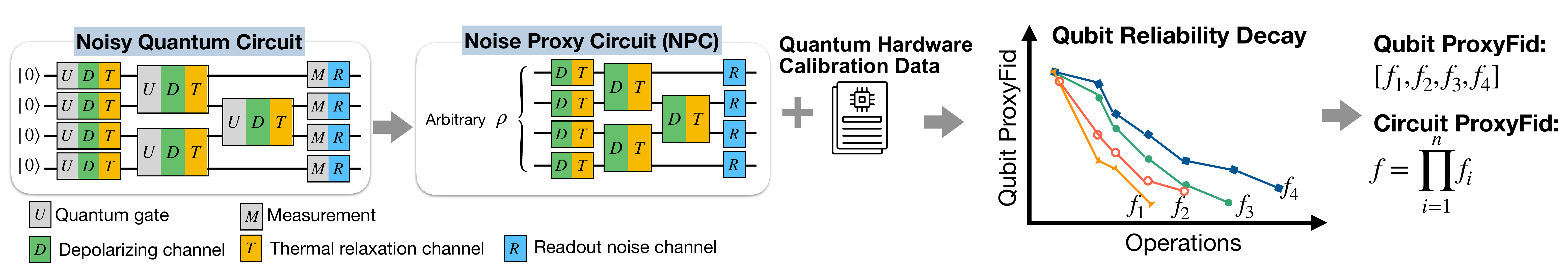}
\caption{Overview of the proposed reliability evaluation framework. Given a noisy quantum circuit, a corresponding NPC is constructed by removing all logical gates while preserving the complete sequence of noise channels, including depolarizing, thermal relaxation, and SPAM errors. Using quantum hardware calibration data, the framework estimates each qubit’s reliability decay through successive noise channels in a state-independent and execution-free manner, and derives a set of qubit proxy fidelities $\{f_1, f_2, \ldots, f_n\}$. These qubit-level reliabilities are then aggregated into the circuit proxy fidelity $f = \prod_{i=1}^{n} f_i$.}
\label{fig:overview}
\end{figure*}

In general, the reliability analysis of quantum circuits consists of two fundamental components. The first is the design of a \textbf{reliability metric} that quantitatively characterizes how reliably a circuit can perform its intended computation. The second is the construction of an \textbf{evaluation model} that estimates the value of this metric based on the circuit structure and the noise characteristics of the quantum hardware.

A widely used metric for circuit reliability is \textit{state fidelity}, which measures the similarity between the noisy quantum state produced by a circuit and its ideal output state. Although fidelity provides an accurate and comprehensive assessment of circuit reliability, its computation is extremely costly, as reconstructing the noisy state typically requires quantum state tomography with a large number of measurements \cite{cramer2010efficient, struchalin2021experimental}. As a result, fidelity estimation becomes impractical and unscalable for large-scale quantum circuits. 
To alleviate this limitation, a variety of approximation- and learning-based approaches have been proposed to predict circuit fidelity from structural and hardware features \cite{saravanan2021test, liu2020reliability, wang2022quest, zlokapa2020deep, vadali2024quantum}. While these methods improve computational efficiency, they often lack interpretability and fail to explicitly model how noise accumulates and interacts during circuit execution, which can lead to predictions that deviate from actual device behavior.

Beyond fidelity-based measures, simpler metrics such as gate count and circuit depth are often used as proxies for circuit reliability. These metrics are straightforward to compute but are noise-unaware and too coarse-grained to accurately reflect true circuit reliability. The success probability, which quantifies the likelihood of producing a correct output, is applicable only to circuits whose ideal behavior corresponds to a small set of basis states.
Furthermore, distribution-similarity-based metrics, which compare noisy and ideal output distributions, require circuit execution and are applicable only to circuits with measurement operations. Overall, existing reliability metrics either achieve high accuracy at prohibitive computational cost or offer improved efficiency at the expense of accuracy, generality, and interpretability.

There remains a lack of a reliability metric that can accurately evaluate circuit reliability across different types of quantum circuits while remaining efficient, execution-free, and interpretable. To address this gap, we propose a reliability evaluation framework that models how noise accumulates throughout circuit execution without reconstructing the output quantum state or requiring circuit execution. An overview of the proposed framework is shown in Fig.~\ref{fig:overview}.
We apply the proposed framework to superconducting quantum computers, as they represent the most widely adopted and well-supported platform in current quantum computing research. Given a noisy quantum circuit, we construct a \textit{\textbf{Noise Proxy Circuit (NPC)}} by removing all logical operations while preserving the complete sequence of noise channels, including depolarizing noise, thermal relaxation effects (amplitude damping and dephasing), as well as state preparation and measurement (SPAM) errors.

The NPC serves as a noise-oriented abstraction of the original circuit, isolating the effects of noise from logical computation and providing a general representation of how noise impacts circuit reliability. Since the behavior of quantum noise may depend on the underlying quantum state, we assume that the NPC is driven by an arbitrary input state, thereby characterizing noise effects over all possible states rather than a specific instance. The similarity between the ideal and noisy states at the output of the NPC defines our reliability metric, termed \textit{\textbf{Proxy Fidelity}}. We further develop an efficient analytical evaluation model to estimate Proxy Fidelity by tracking how reliability degrades along the sequence of noise channels applied to the quantum system.

\paragraph{Contributions.}
To address the limitations of existing reliability evaluation methods for noisy quantum circuits, we make the following contributions:

\begin{itemize}
    \item \textbf{A general reliability evaluation framework.} We propose a unified framework that models how noise accumulates during circuit execution, enabling efficient and interpretable circuit reliability analysis without requiring quantum execution and costly classical computation.
        
    \item \textbf{Noise Proxy Circuit (NPC).} We introduce the concept of the NPC, which removes all logical operations while preserving the sequence and parameters of noise channels. The NPC isolates noise effects from computation and provides a general abstraction of cumulative noise behavior.    
    
    \item \textbf{Proxy Fidelity metric and evaluation model.} We define a new reliability metric, \textit{Proxy Fidelity}, which quantifies the similarity between the ideal and noisy states of the NPC. We further develop an efficient analytical evaluation model that estimates Proxy Fidelity by tracking reliability degradation along noise channels.

    \item \textbf{Comprehensive evaluation.} We evaluate the proposed framework on a diverse set of circuits using both real and simulated quantum computers, demonstrating that Proxy Fidelity closely correlates with state fidelity while significantly reducing computational cost.
    
\end{itemize}

\paragraph{Paper Organization.}
The remainder of this paper is organized as follows.
Sec.~\ref{sec:back} introduces the necessary background and motivation.
Sec.~\ref{sec:method} presents the proposed reliability evaluation framework, including the Noise Proxy Circuit and the Proxy Fidelity metric, along with the corresponding analytical evaluation model.
Sec.~\ref{sec:eval} evaluates the proposed approach on both real and simulated quantum hardware and compares it with existing reliability metrics.
Sec.~\ref{sec:discussion} discusses the limitations and potential impact of the proposed framework.
Sec.~\ref{sec:related} reviews related work, and Sec.~\ref{sec:conc} concludes the paper.

\section{Background and Motivation} \label{sec:back}

\begin{figure*}[t]
\centering
\includegraphics[width=1\linewidth]{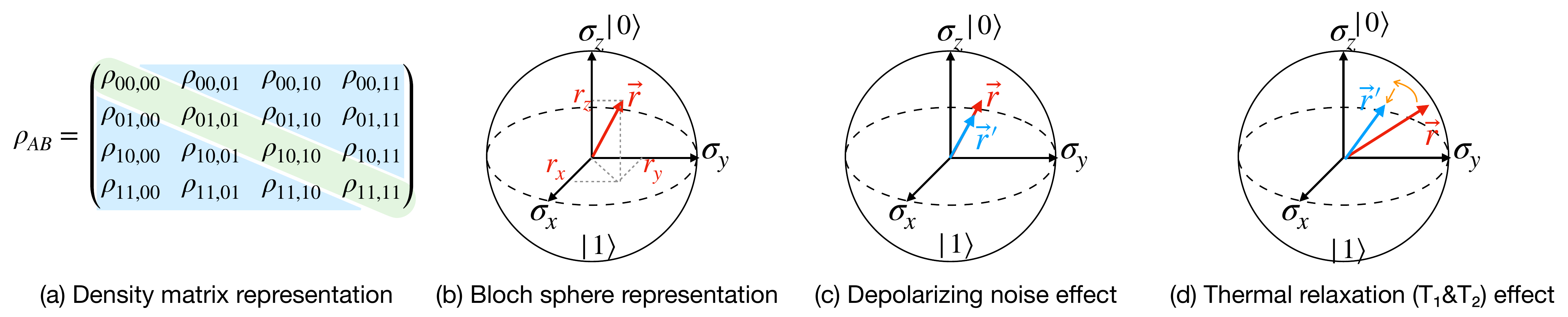}
\caption{Quantum state representations and the effects of common noise channels.}
\label{fig:background}
\end{figure*}

\subsection{Representation of Quantum States}

\subsubsection{Density Matrix}
Quantum computing exploits superposition and entanglement, which enable coherent interference among quantum states and correlations across multiple qubits. As a result, representing quantum information is inherently more complex than in classical systems. To uniformly describe both pure and mixed quantum states, we adopt the \emph{density matrix} formalism. For an $n$-qubit system, the density matrix $\rho$ is a $2^n \times 2^n$ positive semidefinite matrix with unit trace.

For a pure state $|\psi\rangle$, the density matrix is defined as $\rho = |\psi\rangle\langle\psi|$. More generally, a mixed state composed of pure states $\{|\psi_i\rangle\}$ with probabilities $\{p_i\}$ is represented as $\rho = \sum_i p_i |\psi_i\rangle\langle\psi_i|$. This representation naturally captures statistical uncertainty and decoherence, providing a realistic description of quantum systems in noisy environments.
An example of a two-qubit density matrix is illustrated in Fig.~\ref{fig:background}(a). The diagonal elements of $\rho$ correspond to measurement probabilities in the computational basis, while the off-diagonal elements encode quantum coherence. In multi-qubit systems, these off-diagonal terms further capture inter-qubit correlations and reflect the presence of entanglement.

% An example of a two-qubit density matrix is illustrated in Fig.~\ref{fig:background}(a). 
% The \emph{diagonal elements} of $\rho_{AB}$ correspond to the probabilities of measuring the system in each computational basis state. 
% Specifically, $\rho_{00,00}$, $\rho_{01,01}$, $\rho_{10,10}$, and $\rho_{11,11}$ represent the probabilities of obtaining the basis states $|00\rangle$, $|01\rangle$, $|10\rangle$, and $|11\rangle$, respectively, upon measurement. 
% In contrast, the \emph{off-diagonal elements} encode quantum coherence, representing phase correlations between different basis states. 
% In multi-qubit systems, these off-diagonal terms further capture inter-qubit correlations, thereby reflecting the presence of entanglement.

\subsubsection{Bloch Representation}

For a single qubit, the density matrix can be expressed in the \emph{Bloch representation}, which provides a more geometric and intuitive description of quantum states. Any single-qubit density matrix can be written as
\begin{equation}
\rho = \tfrac{1}{2}\left(I + \vec{r}\!\cdot\!\vec{\sigma}\right),
\label{eq:bloch}
\end{equation}
where $\vec{\sigma} = (\sigma_x, \sigma_y, \sigma_z)$ denote the Pauli matrices, which form a basis for representing single-qubit quantum states. In addition, $\vec{r} = (r_x, r_y, r_z)$ is the real-valued \emph{Bloch vector} corresponding to the expectation values of the Pauli operators, given by $r_i = \mathrm{Tr}(\rho \sigma_i)$ for $i \in \{x, y, z\}$. The magnitude of the Bloch vector satisfies $|\vec{r}| \le 1$, where $|\vec{r}| = 1$ represents a pure state, and $|\vec{r}| < 1$ corresponds to a mixed state.

Geometrically, $\vec{r}$ corresponds to a point inside the unit Bloch sphere, as illustrated in Fig.~\ref{fig:background}(b). This representation is particularly useful for analyzing single-qubit noise channels, as common noise processes correspond to simple geometric transformations of the Bloch vector, such as shrinking or rotation. While the Bloch representation is restricted to single-qubit states, it provides valuable intuition for understanding noise-induced state evolution at the physical level.

\subsection{Quantum Noises}
Quantum systems are inherently unstable and highly sensitive to environmental disturbances. In superconducting quantum computers, multiple types of noise coexist, each affecting quantum states in different ways and described by distinct models. The composition of these noises makes the analysis of system reliability particularly challenging.

\subsubsection{Gate Errors (Depolarizing Noise).}

Quantum gate errors are commonly modeled using a \emph{depolarizing channel}, which represents random errors that destroy both population and phase information, leading the system toward the maximally mixed state. Formally, the depolarizing channel is defined as
\begin{equation}
    \mathcal{D}_{\text{depol}}(\rho, p) = (1-p)\rho + p\frac{I}{d},
\end{equation} \label{eq:gateerror}
where $\rho$ is the density matrix of the $n$-qubit quantum state prior to the channel, $d = 2^n$ is the dimension of the quantum system, and $p$ is the depolarizing probability.

In the Bloch representation, the depolarizing channel induces an isotropic contraction of the Bloch vector,
\begin{equation}
\vec{r}' = (1-p)\vec{r}.
\label{eq:dep}
\end{equation}
As illustrated in Fig.~\ref{fig:background}(c), this contraction uniformly reduces the state’s purity while preserving its direction, driving the state toward the maximally mixed state.

\subsubsection{Thermal Relaxation Noise (Amplitude Damping and Dephasing).}
Thermal relaxation is a major source of noise in superconducting quantum processors, modeling energy dissipation and loss of phase coherence over time. 
This process is characterized by two time constants: the amplitude damping time $T_1$ and the dephasing time $T_2$. 
Amplitude damping describes the decay from the excited state $|1\rangle$ to the ground state $|0\rangle$, while dephasing randomizes the relative phase between basis states without changing their populations.

In the Bloch representation, the evolution of a qubit under thermal relaxation during an operation of duration $t$ is given by
\begin{equation}
\label{eq:trc_back}
    r_x' = e^{-t/T_2}r_x, \qquad
r_y' = e^{-t/T_2}r_y, \qquad
r_z' = e^{-t/T_1}r_z + \left(1 - e^{-t/T_1}\right).
\end{equation}
The exponential decay terms $e^{-t/T_1}$ and $e^{-t/T_2}$ correspond to the gradual loss of energy and phase coherence, respectively. 
As a result, the transverse components of the Bloch vector decay due to dephasing, while the longitudinal component relaxes toward the ground state $|0\rangle$ due to amplitude damping. 
Geometrically, thermal relaxation induces an anisotropic contraction of the Bloch sphere accompanied by an upward drift along the $z$-axis, as illustrated in Fig.~\ref{fig:background}(d).

\subsubsection{State Preparation and Measurement Errors (SPAM)}
Errors can also arise during qubit state preparation and measurement. 
For a single qubit, these processes are commonly modeled as a classical bit-flip channel with error rate $e$, where the prepared or measured outcome is incorrect with probability $e$ and correct with probability $1-e$. 
Unlike gate and relaxation noise, SPAM errors do not modify the quantum state during evolution but affect the classical input and output of quantum circuits.

\subsection{Quantum Circuit Reliability Metrics }
In noisy quantum computers, circuit executions inevitably deviate from their ideal behavior due to various noise sources. 
A variety of metrics have been proposed to quantify the reliability of quantum circuits under noise, each capturing different aspects of circuit behavior.

\subsubsection{Gate Count and Circuit Depth.}
Gate count and circuit depth are among the most basic reliability metrics. 
These noise-unaware metrics rely solely on structural properties of a circuit, based on the assumption that shorter circuits experience less accumulated noise. 
However, they fail to accurately reflect actual circuit reliability, as they ignore heterogeneity in noise levels across qubits and gate operations.

\subsubsection{Success Probability.}
The success probability measures how often a noisy circuit produces the expected output bitstring. 
It is derived from the output distribution obtained through circuit execution and is computed as the total probability assigned to the target basis state. 
This metric is applicable only to circuits with a single correct outcome and does not extend to quantum algorithms whose ideal outputs are distributions over multiple basis states.

\subsubsection{Distribution Similarity.}

Circuit reliability can also be evaluated by comparing the output probability distributions of noisy and ideal executions. 
Distribution-similarity metrics, such as the \emph{Hellinger distance}, quantify the statistical divergence between the observed measurement distribution $P_{\mathrm{noisy}}$ and the ideal distribution $P_{\mathrm{ideal}}$:
\[
H(P_{\text{ideal}}, P_{\text{noisy}}) = 
\frac{1}{\sqrt{2}}\sqrt{\sum_i\left(\sqrt{P_{\text{ideal}}(i)} - \sqrt{P_{\text{noisy}}(i)}\right)^2}.
\]
These metrics provide a global statistical comparison across all measurement outcomes. 
However, they operate purely at the classical distribution level, requiring circuit execution and failing to capture quantum properties such as coherence and entanglement.

\subsubsection{State Fidelity.}
Fidelity quantifies the overall similarity between the noisy output state $\rho_{\text{noisy}}$ and the ideal output $\rho_{\text{ideal}}$. 
It is defined as
\[
F(\rho_{\text{ideal}}, \rho_{\text{noisy}}) = 
\left(\mathrm{Tr}\sqrt{\sqrt{\rho_{\text{ideal}}}\rho_{\text{noisy}}\sqrt{\rho_{\text{ideal}}}}\right)^2.
\]
State fidelity takes values in the range $[0,1]$, with $F=1$ indicating identical states and $F=0$ corresponding to orthogonal states. 
When at least one of the two states is pure, the fidelity expression simplifies to $\text{Tr}(\rho_{ideal}\ \rho_{noisy})$, where $\text{Tr}()$ is the trace operation. It can be intuitively understood as the overlap between the ideal and noisy states.

\subsection{Motivation}

\begin{figure*}[t]
\centering
\includegraphics[width=0.8\linewidth]{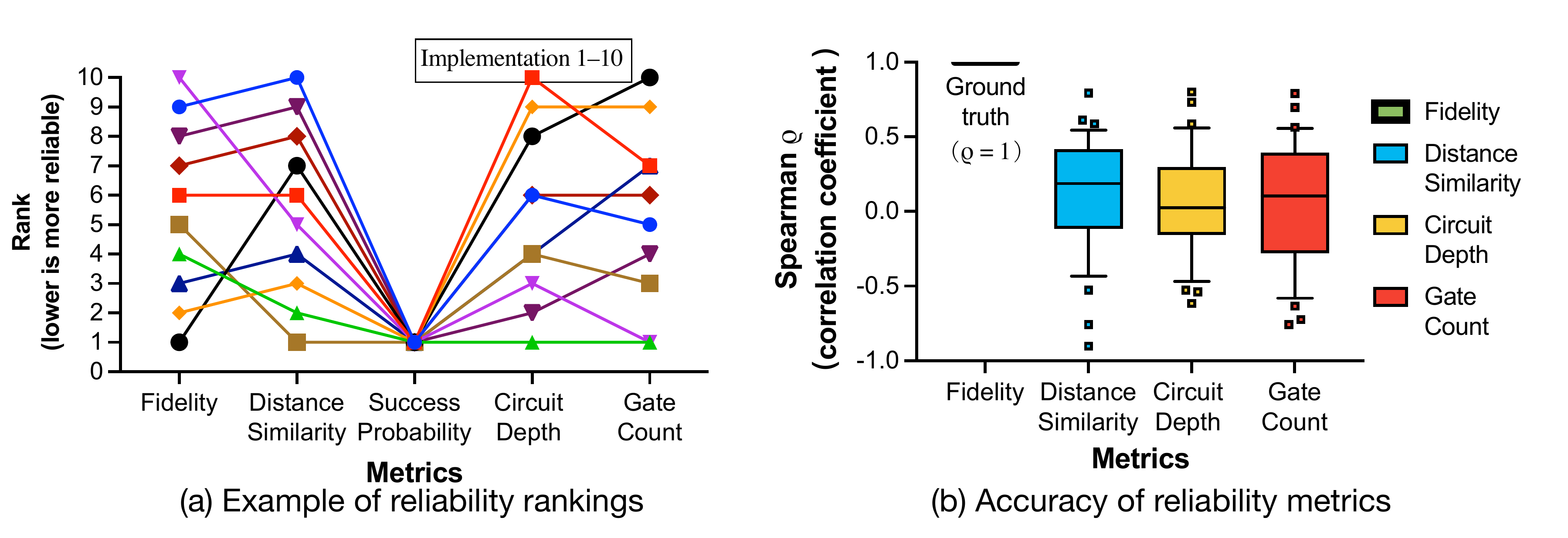}
\caption{
Illustration of the inconsistency and inaccuracy of existing reliability metrics. 
(\textbf{a}) Example of reliability rankings across different metrics for ten compiled implementations of the same random circuit with 10 qubits and 5 layers. 
(\textbf{b}) Consistency of reliability metrics measured by the Spearman correlation coefficient ($\rho$) between each metric’s ranking and the fidelity-based ground truth. 
Non-fidelity metrics exhibit low and unstable correlations, confirming their limited ability to capture true circuit reliability.}
\label{fig:motivation_pair}
\end{figure*}

Although a variety of reliability metrics have been proposed for evaluating noisy quantum circuits, their accuracy and consistency in reflecting actual circuit reliability remain unclear. 
To investigate this issue, we generate a set of random quantum circuits with qubit counts ranging from 10 to 15 and randomly chosen circuit depths between 1 and 10. 
Each circuit is compiled into 10 distinct implementations targeting the noisy backend \texttt{ibm\_torino}, provided by IBM Quantum. 
For each implementation, we evaluate several commonly used reliability metrics, including circuit depth, gate count, success probability, distribution similarity, and fidelity. 
Among these metrics, fidelity is treated as the ground-truth reference for circuit reliability. 
The noisy output states are obtained from a simulator calibrated to the \texttt{ibm\_torino} backend, while the corresponding ideal output states are generated using an ideal quantum simulator.

We rank the 10 implementations of each circuit according to each metric and compare these rankings against the fidelity-based ground truth. 
Fig.~\ref{fig:motivation_pair}(a) illustrates an example of the reliability rankings produced by different metrics for a single circuit. 
The rankings vary significantly across metrics, demonstrating the \textit{inconsistency among these metrics in reflecting circuit reliability}.
To quantify this inconsistency, Fig.~\ref{fig:motivation_pair}(b) reports the Spearman’s $\rho$ correlation between each metric’s ranking and the fidelity-based ranking across 60 random circuits. 
Although the distribution-similarity metric exhibits relatively higher correlation with fidelity, it still fails to consistently identify the most reliable implementations. 
Overall, all non-fidelity metrics exhibit low and highly variable correlations (typically $\rho < 0.4$), indicating that their rankings are poorly aligned with actual circuit reliability.

These observations reveal a fundamental limitation of existing reliability metrics. 
While fidelity provides the most accurate and general measure of circuit reliability, directly computing fidelity is computationally expensive and experimentally impractical for large circuits, as it requires quantum state tomography or repeated hardware execution. 
In contrast, structural metrics such as circuit depth and gate count are computationally efficient but overly coarse and insensitive to the underlying noise behavior. 
This trade-off between \textit{accuracy} and \textit{efficiency} exposes a critical gap in current reliability evaluation methods.

Motivated by this gap, we seek a reliability evaluation framework that achieves fidelity-level accuracy in ranking circuit implementations while remaining execution-free and computationally efficient. 
Our key insight is to model how noise accumulates throughout circuit execution, rather than attempting to reconstruct the full quantum state.

\section{Method} \label{sec:method}

\subsection{Overview}
We propose a reliability evaluation framework that is accurate, efficient, and interpretable by explicitly modeling how noise accumulates during circuit execution. 
The framework avoids quantum execution and full quantum state reconstruction, while remaining aligned with fidelity-level reliability estimation. 
Fig.~\ref{fig:overview} illustrates the overall workflow.
Given a noisy quantum circuit, we first construct its corresponding \emph{Noise Proxy Circuit (NPC)} by removing all logical gate operations while preserving the complete sequence of physical noise channels, including depolarizing noise, thermal relaxation, and state preparation and measurement (SPAM) errors. 
The NPC serves as a noise-oriented abstraction of the original circuit, isolating hardware-induced noise effects and capturing how noise propagates and accumulates throughout execution.
Using quantum hardware calibration data, the framework models the reliability decay of each qubit as it passes through successive noise channels in the NPC, yielding a set of qubit-level proxy fidelities $[f_1, f_2, \ldots, f_n]$. 
These qubit proxy fidelities are then aggregated to compute a circuit-level proxy fidelity, which reflects the overall reliability of the original circuit.

\subsection{Reliability Metric Formulation} \label{sec:metric}

\subsubsection{Noise Proxy Circuit}
The reliability of a quantum circuit is determined by how noise accumulates throughout its execution. 
Directly simulating noisy state evolution to capture these effects is computationally prohibitive for large-scale circuits. 
To efficiently characterize noise accumulation while avoiding explicit state simulation, we introduce the \emph{Noise Proxy Circuit (NPC)}, which abstracts the cumulative effect of hardware noise in a given circuit.

Formally, a noisy quantum circuit can be represented as a sequence of alternating ideal operations $\mathcal{U}_i$ and corresponding noise channels $\mathcal{N}_i$:
\[
\mathcal{C}_{\mathrm{ori}} = \mathcal{N}_L \circ \mathcal{U}_L \circ \mathcal{N}_{L-1} \circ \mathcal{U}_{L-1} \circ \cdots \circ \mathcal{N}_1 \circ \mathcal{U}_1.
\]
Its NPC is obtained by removing all operations while preserving the complete sequence of noise channels:
\[
\mathcal{C}_{\mathrm{NPC}} = \mathcal{N}_L \circ \mathcal{N}_{L-1} \circ \cdots \circ \mathcal{N}_1.
\]
Each $\mathcal{N}_i$ may include a depolarizing channel $\mathcal{D}$ and a thermal relaxation channel $\mathcal{T}$ associated with quantum gates, or a bit-flip channel $\mathcal{R}$ for SPAM. These channels are parameterized by the hardware calibration data of the target quantum processor. While Fig.~\ref{fig:overview} depicts noise channels as applied after each quantum operation for clarity, they act concurrently with the corresponding operations during execution.

To eliminate state-dependent effects and capture intrinsic noise accumulation, we model the NPC as acting on an arbitrary input state and focus on state-agnostic properties of the resulting noise evolution. 
This abstraction enables a circuit-structure-driven characterization of cumulative noise effects, which forms the foundation for defining our reliability metric.

\subsubsection{Proxy Fidelity}
We define a new reliability metric, termed \emph{Proxy Fidelity}, to quantify circuit reliability degradation induced by the NPC. 
Proxy fidelity measures the similarity between an input quantum state $\rho$ and the corresponding output state $\rho'$ after applying the NPC, where a higher value indicates higher circuit reliability. 
To enable fine-grained and interpretable evaluation, proxy fidelity is defined hierarchically at both the qubit level and the circuit level.

We assume that noise channels act independently and affect only the qubits to which they are applied. 
Under this assumption, the reliability of each qubit decays progressively along the sequence of noise channels it experiences within the NPC. 
The proxy fidelity of the $i$-th qubit, denoted $f_i$, is defined as
\begin{equation}
    f_i = \mathrm{Tr}\big(\rho_i \, \mathcal{C}_{\mathrm{NPC}}^{(i)}(\rho_i)\big),
\end{equation}
where $\mathcal{C}_{\mathrm{NPC}}^{(i)}$ denotes the composition of noise channels acting on qubit $i$, and $\rho_i$ is an arbitrary single-qubit input state. 
By considering an arbitrary input state, $f_i$ characterizes state-agnostic reliability degradation induced by noise, rather than behavior tied to a specific circuit input.

The trace (Hilbert-Schmidt) inner product captures the similarity between two quantum states by accounting for both population distributions (diagonal elements) and quantum coherence (off-diagonal elements). 
Although it is not equivalent to full quantum state fidelity, this measure preserves monotonic alignment with fidelity under common noise channels, while remaining significantly more efficient to compute. 
It therefore serves as a practical and well-aligned proxy for fidelity in reliability evaluation.

The overall circuit-level proxy fidelity is obtained by aggregating all $n$ qubit-level proxy fidelities:
\begin{equation}
    f = \prod_{i=1}^{n} f_i.
    \label{eq:circ_fid}
\end{equation} 
When the ideal circuit output is a single computational basis state, the final quantum state is separable and can be written as $\rho = \bigotimes_i \rho_i$. 
In this case, the multiplicative aggregation exactly composes the reliability of the circuit from the reliabilities of individual qubits.

More generally, when entanglement is present in the output state, direct evaluation of circuit-level reliability requires full-state simulation or execution. 
We therefore adopt the multiplicative aggregation as a state-independent and scalable approximation. 
This formulation captures how noise-induced reliability degradation accumulates along individual qubit trajectories and preserves the monotonic property that overall circuit reliability decreases whenever any constituent qubit becomes unreliable. 
As demonstrated in Fig.~\ref{fig:eval_1_2}, this approximation is effective in practice for estimating circuit reliability across diverse circuits.

\begin{figure*}[t]
\centering
\includegraphics[width=0.6\linewidth]{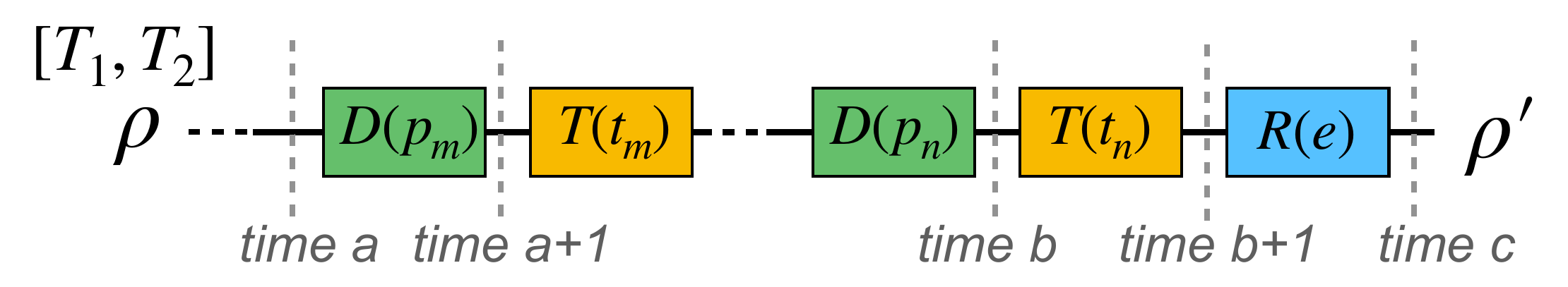}
\caption{Schematic illustration of the qubit Noise Proxy Circuit.
The qubit evolves from the initial state $\rho$ to $\rho'$ through depolarizing $\mathcal{D}(p)$, thermal relaxation $\mathcal{T}(t)$, and SPAM noise $\mathcal{R}(e)$ channels.
The final proxy fidelity is computed according to Eq.~\ref{eq:qubit_fid}.
}
\label{fig:method_1}
\end{figure*}

\subsection{Qubit-Level Reliability Evolution}
We now analyze how different noise channels degrade the reliability of a single qubit and how the corresponding proxy fidelity can be evaluated.
To illustrate this process, Fig.~\ref{fig:method_1} presents a schematic example where a qubit evolves under sequential noise channels.

Each qubit is characterized by its SPAM error rates, an amplitude damping time $T_1$, and a dephasing time $T_2$.
During execution, the qubit is subject to a sequence of depolarizing channels $\mathcal{D}(p)$ with depolarizing probability $p$, thermal relaxation channels $\mathcal{T}(t)$ parameterized by the operation duration $t$, which induce both amplitude damping ($T_1$) and phase damping ($T_2$) effects, and a final SPAM noise channel $\mathcal{R}(e)$ with bit-flip rate $e$.

The qubit starts in an arbitrary initial state $\rho = \frac{1}{2}( I + \vec{r} \cdot \vec{\sigma})$, with an initial \textit{state-independent} proxy fidelity $f_0 = 1$, indicating that the state has not yet been affected by any noise process.
The final proxy fidelity of the qubit after passing through all noise channels is given by the following closed-form expression:
\begin{equation}
    f = \Big[\frac{1}{2} + \Big(f_0 - \frac{1}{2}\Big) \prod_{i=1}^{n}(1 - p_i)
    \Big(\tfrac{2}{3}e^{-t_i/T_2} + \tfrac{1}{3}e^{-t_i/T_1}\Big)\Big](1 - e)
    \label{eq:qubit_fid}
\end{equation}
where $n$ denotes the number of depolarizing–thermal relaxation channel pairs experienced by the qubit.

We next detail how each type of noise channel incrementally updates the qubit’s proxy fidelity, leading to the above expression. 
For clarity, we denote the qubit proxy fidelity after the $i$-th noise update as $f_i$, where the index $i$ enumerates the evolution steps of a single qubit. This indexing is distinct from Sec.~\ref{sec:metric}, in which $i$ indexes different qubits.

\subsubsection{Depolarizing Channel}
We illustrate the impact of depolarizing noise on qubit reliability using the transition between time steps $a$ and $a{+}1$ in Fig.~\ref{fig:method_1}. 
At time $a$, the qubit proxy fidelity is defined as $f_{a} = \mathrm{Tr}(\rho\, \rho_{a})$, where $\rho$ denotes the reference input state and $\rho_a$ is the current noisy state of the qubit. 
$\rho_a$ can be expressed as
\[
\rho_{a} = \frac{1}{2}\!\left[I + (2f_{a} - 1)\, \vec{r}\!\cdot\!\vec{\sigma}\right].
\]
At time $a+1$, after the qubit passes through the depolarizing channel $\mathcal{D}(p_m)$ as defined in Eq.~\ref{eq:dep}, its state evolves to
\[
\rho_{a+1} = \frac{1}{2}\!\left[I + (2f_{a} - 1)(1 - p_m)\, \vec{r}\!\cdot\!\vec{\sigma}\right].
\]
The updated proxy fidelity at time $a{+}1$ is therefore given by
\begin{equation}
f_{a+1} 
= \mathrm{Tr}(\rho\, \rho_{a+1}) 
= \frac{1}{2} + \left(f_{a} - \frac{1}{2}\right)(1 - p_m).
\label{eq:proxyf_dep_iter}
\end{equation}
Eq.~\ref{eq:proxyf_dep_iter} quantifies how the depolarizing channel linearly reduces the qubit reliability as a function of the depolarizing probability $p_m$ and previous proxy fidelity $f_a$.
This degradation depends solely on the channel parameter and is independent of the specific input state.
When $p_m = 0$, the channel introduces no noise, resulting in $f_{a+1} = f_a$, meaning the qubit reliability is fully preserved.
Conversely, when $p_m = 1$, the depolarizing channel completely randomizes the qubit and erases all information encoded in the qubit, producing a maximally mixed state $\rho_{a+1} = I/2$, and the proxy fidelity drops to $f_{a+1} = \mathrm{Tr}(\rho\ I/2) = 1/2$. \\

\begin{figure*}[t]
\centering
\includegraphics[width=1\linewidth]{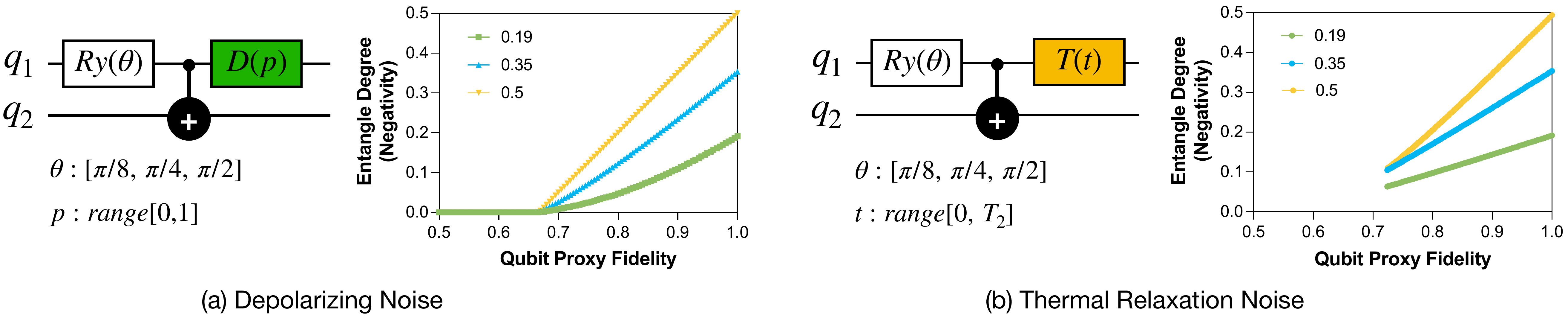}
\caption{Relationship between qubit proxy fidelity and entanglement degree (negativity) under different noise channels. For both depolarizing (a) and thermal relaxation (b) noise, the entanglement degree decreases as noise strength increases, and the estimated qubit proxy fidelity decreases accordingly.
The strong proportional relationship between the two confirms that the proxy fidelity effectively reflects noise-induced degradation of quantum correlations.
The numerical procedure is provided in Appendix \ref{app:fig_5}.}
\label{fig:method_2}
\end{figure*}

It is worth noting that the initial state of a qubit in this framework may be either pure or mixed, depending on its entanglement with other qubits in the overall quantum system.
We initialize the reliability evolution with a normalized proxy fidelity of $f_0 = 1$, corresponding to the state before any noise is applied. 
When a qubit is entangled with others, local depolarizing noise reduces the strength of entanglement, which is reflected by a decrease in the qubit proxy fidelity.

As illustrated in Fig.~\ref{fig:method_2} (a), we prepare a two-qubit system in which qubit $q_1$ is rotated by a single-qubit gate $R_y(\theta)$ with $\theta \in \{\pi/8, \pi/4, \pi/2\}$ to generate different initial entanglement strengths with qubit $q_2$ through a subsequent CNOT operation. A depolarizing channel $\mathcal{D}(p)$ with parameter $p \in [0,1]$ is then applied to $q_1$ to add depolarized noise.
As $p$ increases, the entanglement degree (measured by negativity) monotonically decreases, and the corresponding qubit proxy fidelity computed from Eq.~\ref{eq:proxyf_dep_iter} decreases accordingly.
The monotonic relationship between entanglement reduction and proxy fidelity degradation shows that proxy fidelity, which is initialized to 1, effectively captures the extent of entanglement loss in a qubit, thereby serving as a consistent and interpretable measure of qubit reliability.

\subsubsection{Thermal Relaxation Channel}

We next quantify how the thermal relaxation channel contributes to the degradation of a qubit’s proxy fidelity by describing the evolution from time $b$ to time $b{+}1$ in Fig.~\ref{fig:method_1}, during which the qubit undergoes a relaxation process of duration $t_n$.

Let the qubit state at time $b$ be
\[
\rho_b = \tfrac{1}{2}\!\left(I + \vec{s}\!\cdot\!\vec{\sigma}\right), \qquad \vec{s} = (s_x, s_y, s_z)
\]
After passing through a thermal relaxation channel, the qubit experiences both amplitude damping and phase damping, and its state evolves to
\[
\rho_{b+1} = \tfrac{1}{2}\!\left(I + \vec{s}'\!\cdot\!\vec{\sigma}\right),
\qquad
\vec{s}' = \big(e^{-t_n/T_2}s_x,\; e^{-t_n/T_2}s_y,\; e^{-t_n/T_1}s_z + (1 - e^{-t_n/T_1})\big),
\]
where $T_1$ and $T_2$ denote the amplitude damping and dephasing times, respectively.
The current proxy fidelity is $f_{b+1} = \mathrm{Tr}(\rho\, \rho_{b+1}) = \tfrac{1}{2}\!\left(1 + \vec{r}\!\cdot\!\vec{s}'\right)$, where $\rho = \tfrac{1}{2}(I + \vec{r}\!\cdot\!\vec{\sigma})$ is the initial state. 
Similarly, the proxy fidelity at time $b$ is $f_b = \mathrm{Tr}(\rho\,\rho_b) = \tfrac{1}{2}(1 + \vec{r}\!\cdot\!\vec{s})$, which implies $\vec{r}\!\cdot\!\vec{s} = 2f_b - 1$.  
Substituting this relation yields
\[
\begin{aligned}
\vec{r}\!\cdot\!\vec{s}' 
&= e^{-t_n/T_2}(r_x s_x + r_y s_y) + r_z\!\big(e^{-t_n/T_1}s_z + 1 - e^{-t_n/T_1}\big) \\
% &= e^{-t_n/T_2}(\vec{r}\!\cdot\!\vec{s} - r_z s_z) + e^{-t_n/T_1}r_z s_z + r_z(1 - e^{-t_n/T_1}) \\
&= e^{-t_n/T_2}(2f_b - 1) + (e^{-t_n/T_1} - e^{-t_n/T_2})\,r_z s_z + r_z(1 - e^{-t_n/T_1}).
\end{aligned}
\]
Thus, the proxy fidelity after the thermal relaxation channel is
\[
f_{b+1} = \tfrac{1}{2}\!\left[1 + e^{-t_n/T_2}(2f_b - 1) + (e^{-t_n/T_1} - e^{-t_n/T_2})\,r_z s_z + r_z(1 - e^{-t_n/T_1})\right].
\]
This expression shows that the reliability degradation induced by thermal relaxation is inherently \textit{state-dependent}.

To obtain a state-agnostic update rule, we adopt an isotropic and unbiased prior for the unknown initial direction of $\vec{r}$, under which $\mathbb{E}[\vec{r}] = \mathbf{0}$ and $\mathbb{E}[\vec{r}\,\vec{r}^{\!\top}] = \tfrac{\alpha}{3}I$ with $\alpha = \mathbb{E}[\|\vec{r}\|^2] \in [0, 1]$. 
Then $\mathbb{E}[r_z] = 0$ and $\mathbb{E}[r_z s_z] = \tfrac{1}{3}\,\mathbb{E}[\vec{r}\!\cdot\!\vec{s}]$. 
The expected proxy fidelity update therefore becomes
\begin{equation}
    f_{b+1} = 
\tfrac{1}{2} + \Big(f_b - \tfrac{1}{2}\Big)
\Big(\tfrac{2}{3}e^{-t_n/T_2} + \tfrac{1}{3}e^{-t_n/T_1}\Big).
\label{eq:trc}
\end{equation} 
This isotropic averaging yields a \textit{state-independent} proxy fidelity update that effectively captures the average effect of amplitude and phase damping over all possible initial states.

Similar to the depolarizing noise, the thermal relaxation channel also reduces the correlations between entangled qubits. As shown in Fig.~\ref{fig:method_2} (b), we prepare the same two-qubit configuration, where qubit $q_1$ is rotated by $R_y(\theta)$ with $\theta \in \{\pi/8, \pi/4, \pi/2\}$ to generate different initial entanglement strengths with qubit $q_2$ through a subsequent CNOT operation.
A thermal relaxation channel $\mathcal{T}(t)$ with duration $t \in [0, T_2]$ is then applied to $q_1$.
As the relaxation duration increases, both the entanglement degree (negativity) and the qubit proxy fidelity obtained from Eq.~\ref{eq:trc} decrease monotonically.
This consistent trend further confirms that the proxy fidelity effectively tracks the reliability degradation induced by relaxation-related entanglement loss, reinforcing its general applicability across different noise mechanisms.

\subsubsection{SPAM Noise Channel}
The SPAM noise channel is modeled as a classical bit-flip process with an error rate $e$, which flips the measured outcome with probability $e$ in NPC and is independent of state.
Importantly, this process does \emph{not} alter the quantum state prior to measurement; instead, it perturbs only the observed classical outcome distribution.
Let $f_{c-1}$ denote the proxy fidelity of the qubit immediately before measurement.
The final proxy fidelity is therefore given by $f_c = f_{c-1} \times (1-e)$.

\subsection{Circuit-Level Reliability Integration}
Building on the qubit-level reliability evolution described above, we now extend the proxy fidelity evaluation to the entire quantum circuit.
Our approach estimates circuit reliability by systematically aggregating qubit-level proxy fidelities while respecting the structural dependencies imposed by the original circuit.
In particular, we identify three key factors that influence circuit-level reliability estimation, which we discuss in detail below.

\begin{figure*}[t]
\centering
\includegraphics[width=1\linewidth]{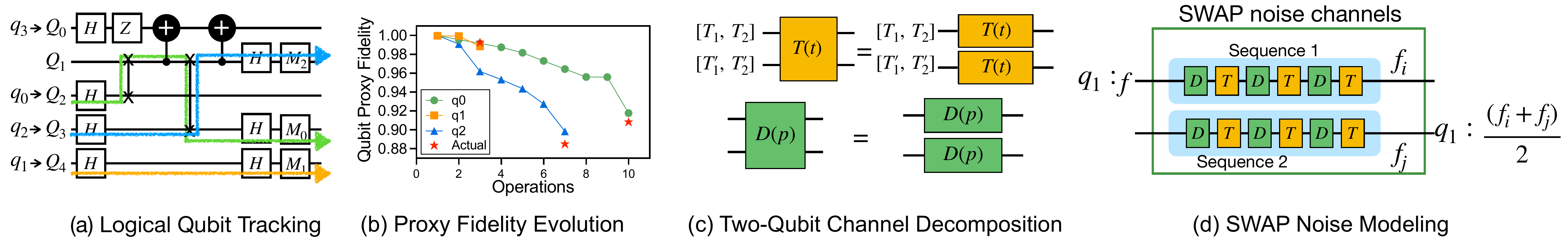}
\caption{Illustration of circuit-level proxy fidelity evaluation.
(a) Tracking reliability along logical qubit trajectories during circuit execution.
(b) Qubit proxy fidelity decay across operations, compared with actual fidelity.
(c) Decomposition of two-qubit noise channels into parallel single-qubit channels.
(d) Modeling SWAP-induced noise via two channel sequences.
}
\label{fig:method_3}
\end{figure*}

\subsubsection{Logical Qubit Tracking}
Circuit-level proxy fidelity is computed by applying the qubit-level update rule in Eq.~\ref{eq:qubit_fid} to each \emph{logical qubit} and aggregating the resulting qubit-level proxy fidelities.
Importantly, we track reliability along the sequence of noise channels experienced by each logical qubit, rather than by individual physical qubits, as illustrated in Fig.~\ref{fig:method_3}(a).
This design reflects the fact that the reliability of a logical qubit degrades cumulatively due to the noisy operations applied to it, even though these operations may be executed on different physical qubits.

During circuit compilation, logical qubits are initially mapped to physical qubits. However, this mapping may change dynamically as logical qubits are routed via SWAP operations to enable two-qubit gates between non-adjacent qubits.
As a result, a logical qubit may traverse multiple physical locations while accumulating noise from the sequence of operations along its execution path.
Because noise parameters such as depolarizing probability $p$, relaxation times $(T_1, T_2)$, and operation duration $t$ depend on the qubit’s \emph{current} physical residence, the routing process determines the exact sequence of noise channels encountered by each logical qubit.
We therefore analyze the compiled circuit to reconstruct the trajectory of each logical qubit and apply the corresponding noise updates along its path. Fig.~\ref{fig:method_3}(a) highlights the trajectories of logical qubits in an example compiled circuit, while Fig.~\ref{fig:method_3}(b) shows the temporal evolution of qubit-level proxy fidelity along these trajectories.
As logical qubits undergo successive operations, their proxy fidelities decrease progressively, reflecting the cumulative impact of hardware noise.
Moreover, the distinct degradation patterns observed across qubits reveal the heterogeneous effects of different noise channels on overall circuit reliability.

\subsubsection{Two-Qubit Noise Channel Decomposition}
A two-qubit gate acting on a pair of logical qubits introduces two sources of noise:
(1) gate-time thermal relaxation on each participating qubit, and (2) a two-qubit depolarizing channel.
For individual qubit reliability analysis, we decompose these channels into two parallel single-qubit channels applied independently to the two logical qubits involved, as shown in Fig.~\ref{fig:method_3} (c).

During a two-qubit gate of duration $t$, each participating logical qubit experiences a local thermal relaxation channel parameterized by the $T_1$ and $T_2$ values of its current physical qubit.

In addition, a two-qubit depolarizing channel with parameter $p$ can be approximated by two parallel single-qubit depolarizing channels. Each channel independently affects its corresponding qubit with the same depolarizing probability $p$, as illustrated in Fig.~\ref{fig:method_3} (c).
Formally, consider a two-qubit state represented by the density matrix \( \rho_{AB} \). The state of the individual qubits are \( \rho_A = \text{Tr}_B(\rho_{AB}) \) and \( \rho_B = \text{Tr}_A(\rho_{AB}) \), respectively. After applying a 2-qubit depolarizing channel with parameter \( p \) to \( \rho_{AB} \), the resulting state is \( \rho_{AB}' = \mathcal{D}(\rho_{AB}, p) \). The depolarized states of the qubits are \( \rho_A' = \text{Tr}_B(\rho_{AB}') \) and \( \rho_B' = \text{Tr}_A(\rho_{AB}') \). These states are equivalent to each qubit passing through a single-qubit depolarizing channel with parameter \( p \), such that \( \rho_A' = \mathcal{D}(\rho_A, p) \) and \( \rho_B' = \mathcal{D}(\rho_B, p) \). A detailed derivation is provided in Appendix~\ref{app:2q_dep}.

\subsubsection{SWAP Operation Modeling}
A special case in circuit-level reliability analysis is the SWAP operation, which exchanges the physical locations of two logical qubits $q_0$ and $q_1$.
A SWAP gate is logically implemented using three CNOT gates:
\begin{equation*}
\label{eq:swap}
\text{SWAP}(q_0, q_1)
=
\text{CNOT}(q_0, q_1)\,
\text{CNOT}(q_1, q_0)\,
\text{CNOT}(q_0, q_1),
\end{equation*}
During circuit compilation, each CNOT gate is further decomposed into a sequence of native single- and two-qubit gates supported by the target quantum processor.
For example, the IBM quantum processor \texttt{ibm\_torino} supports the native gate set
\([\texttt{cz}, \texttt{id}, \texttt{rx}, \texttt{rz}, \texttt{rzz}, \texttt{sx}, \texttt{x}]\).
Under this gate set, one possible realization of a \(\mathrm{CNOT}(q_1, q_2)\) operation is
\[
\mathrm{CNOT}(q_1, q_2)
=
\mathrm{RZ}_{q_2}(\pi)\;
\mathrm{SX}_{q_2}\;
\mathrm{RZ}_{q_2}(\pi)\;
\mathrm{CZ}(q_1, q_2)\;
\mathrm{RZ}_{q_2}(\pi)\;
\mathrm{SX}_{q_2}\;
\mathrm{RZ}_{q_2}(\pi).
\]
During the execution of this sequence, the two logical qubits are temporarily interleaved across both physical qubits before being fully exchanged.
Rather than modeling each intermediate relocation explicitly, we treat the SWAP operation as a single integrated noise segment.
Within this segment, each involved physical qubit experiences its corresponding sequence of depolarizing and thermal relaxation channels induced by the native gates.
To compute the post-SWAP reliability of a logical qubit, we aggregate the noise effects accumulated on both physical qubits.
Specifically, as illustrated in Fig.~\ref{fig:method_3}(d), suppose a logical qubit has a pre-SWAP proxy fidelity $f$.
The noise-channel sequences on the two physical qubits update this fidelity to $f_i$ and $f_j$, respectively.
After the SWAP operation transfers the logical qubit to the other physical location, its post-SWAP proxy fidelity is given by $(f_i + f_j) / 2$.
This averaging-based integration provides a conservative and efficient approximation of reliability degradation during SWAP operations.

\subsection{Complexity Analysis}
The proposed circuit reliability evaluation method estimates qubit-level proxy fidelities based on the circuit structure and the noise parameters associated with each noise channel. Unlike state-fidelity estimation, which requires exponential-time quantum state tomography, our approach performs purely classical computations that scale linearly with circuit size.

Considering a compiled circuit composed of $k$ logical qubits, $n$ single-qubit operations, and $m$ two-qubit operations, with at most $k$ qubit measurements. 
Each single-qubit operation updates the proxy fidelity of the corresponding qubit, while each two-qubit operation affects both involved qubits, with each gate inducing both depolarizing channel and thermal relaxation channels.
Therefore, the total computational complexity for evaluating all qubits is \( O(n + m + k) \), which scales linearly with the number of circuit operations.
In summary, the proposed method achieves linear-time complexity with respect to circuit size, making it highly efficient and scalable compared with full density-matrix simulation, measurement-based fidelity estimation methods, and ML-based methods.

\section{Evaluation} \label{sec:eval}

We evaluate the proposed reliability evaluation framework from the perspectives of accuracy, efficiency, and interpretability on both real and simulated noisy quantum computers.
All experiments are conducted using the Qiskit Python library, executed either on IBM Quantum Cloud hardware or on noise-aware simulators.

The real quantum devices used in our experiments include \texttt{ibm\_torino} (133 qubits), \texttt{ibm\_osaka} (127 qubits), \texttt{ibm\_brisbane} (127 qubits), and the retired \texttt{ibmq\_perth} (7 qubits).
The simulated noisy quantum computers are constructed using the corresponding device noise models.
We consider five representative classes of quantum circuits: Bernstein--Vazirani (BV) circuits, Greenberger--Horne--Zeilinger (GHZ) circuits, identity (ID) circuits, random circuits, and variational quantum circuits (VQCs).
All circuits are compiled using Qiskit's transpiler with optimization level 0.

Hardware calibration data are obtained from the IBM Quantum backends at the time of circuit execution. These data provide the parameters of noisy operations and qubits, including gate error rates, qubit relaxation and dephasing times, gate durations, and readout error probabilities.
For each quantum gate, we derive the depolarizing parameter $p$ from the reported error rate $r$ using the relation $p = 1 - r d/(d-1)$, where $d$ denotes the dimension of the gate operation~\cite{magesan2012characterizing}.
Thermal relaxation noise is modeled using the reported qubit relaxation and dephasing times together with gate durations, while readout noise is directly modeled using the reported readout error rates.

We first evaluate the accuracy of the proposed Proxy Fidelity at both the qubit and circuit levels by comparing it against state fidelity.
We then examine the ranking consistency of proxy fidelity across multiple compiled implementations of the same logical circuit and compare its performance with commonly used reliability metrics and fidelity estimators.
Next, we validate the state-independence of proxy fidelity using VQCs, where circuits with identical structures but different parameter values generate distinct quantum states.
Finally, we present a case study demonstrating how qubit-level reliability analysis provides fine-grained insights into noise accumulation and circuit reliability that are not captured by existing metrics.

\subsection{Fidelity Estimation accuracy}

\begin{figure*}[t]
\centering
\includegraphics[width=1\linewidth]{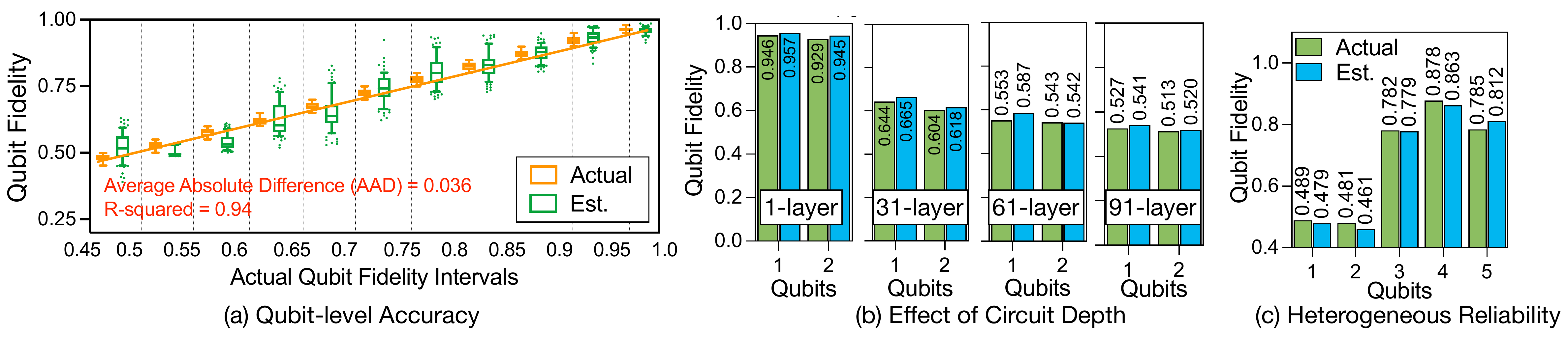}
\caption{Qubit-level fidelity estimation results. 
(a) Overall estimation accuracy across 3,276 readout qubits collected from 320 noisy circuit executions.
The estimated proxy fidelities closely match the ground-truth fidelities, achieving an AAD of 0.036 and a coefficient of determination ($R^2$) of 0.94.
(b) Fidelity degradation with circuit depth in two-qubit ID circuits. As the number of gate layers increases, both actual and estimated fidelities decrease consistently, demonstrating that the proposed method accurately captures cumulative noise effects.
(c) Heterogeneous qubit reliability within a 5-qubit BV circuit. The estimated fidelities closely track per-qubit variations caused by different noise channels, confirming the method’s ability to capture structure-dependent noise accumulation.
}
\label{fig:eval_1_1}
\end{figure*}

\subsubsection{Qubit-Level Evaluation}
To evaluate the qubit-level estimation accuracy, we use BV and ID circuits executed on real IBM quantum hardware.
Each of these circuits produces a single ideal deterministic basis state, allowing the ground-truth fidelity of each qubit to be directly obtained from the measured outcome probabilities, without requiring quantum state tomography or density-matrix simulation.

Fig.~\ref{fig:eval_1_1} (a) compares the estimated qubit proxy fidelities with the actual fidelities across 3,276 readout qubits collected from 320 noisy circuit executions on IBM quantum hardware. The dataset includes 230 BV circuits (ranging from 4 to 30 qubits) and 90 ID circuits (ranging from 4 to 6 qubits, each containing 1–121 random identity layers). The x-axis represents actual qubit fidelity intervals from 0.45 to 1.00, with a step size of 0.05, covering 96.3\% of all qubits. Box plots visualize the distributions of both actual and estimated fidelities within each interval. The proposed proxy fidelity exhibits strong agreement with the true fidelities, achieving an average absolute difference (AAD) of 0.036 and a coefficient of determination ($R^2$) of 0.94.
These results confirm that the proxy fidelity accurately captures qubit-level reliability and serves as a robust proxy for true state fidelity in realistic noisy environments.

Beyond overall accuracy, we further evaluate the robustness of the proposed method under varying levels of noise accumulation and across qubits within the same circuit.
Fig.~\ref{fig:eval_1_1} (b) shows the comparison between the estimated and actual qubit fidelities in two-qubit ID circuits with different numbers of gate layers. In each circuit, both qubits experience the same number of gate layers. As the number of layers increases from 1 to 91, the actual fidelities of both qubits decrease due to cumulative noise, and the estimated fidelities follow the same trend. This close correspondence demonstrates that the proposed method reliably captures fidelity degradation caused by sequential noisy operations.

Moreover, Fig.~\ref{fig:eval_1_1} (c) presents qubit fidelities for a 5-qubit BV circuit. The actual fidelities of different qubits within the same circuit vary significantly because each qubit undergoes a distinct number of operations and experiences different noise channels. Despite this heterogeneity, the estimated proxy fidelities closely match the ground-truth fidelities, indicating that the proposed method accurately captures noise accumulation along individual qubit trajectories.

Overall, these results verify that the proposed proxy fidelity provides both accurate and interpretable qubit-level reliability estimation, forming the foundation for circuit-level analysis presented next.

\begin{figure*}[t]
\centering
\includegraphics[width=1\linewidth]{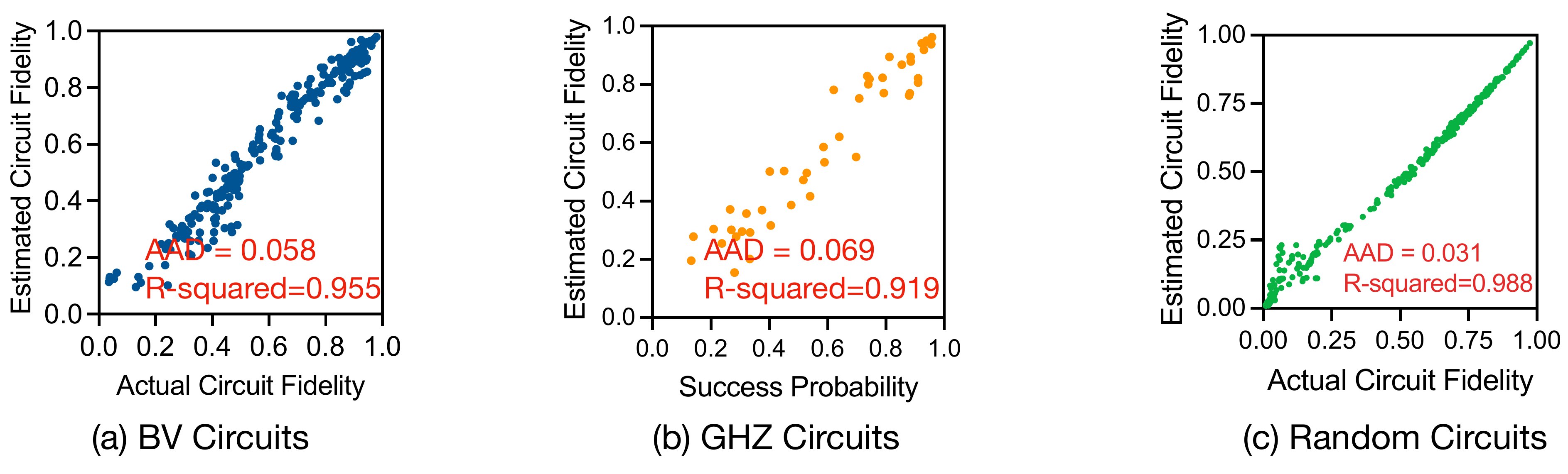}
\caption{
Circuit-level fidelity estimation accuracy for different circuit types.
(a) BV circuits with deterministic outputs.
(b) GHZ circuits generating maximally entangled states.
(c) Random circuits simulated with real-device noise models.
The proposed proxy fidelity consistently approximates true circuit fidelity across all cases.
}
\label{fig:eval_1_2}
\end{figure*}

\subsubsection{Circuit-Level Evaluation}
Building upon the verified qubit-level fidelity estimation, we now evaluate the accuracy of the proposed method at the circuit level. The circuit-level proxy fidelity is computed by aggregating the qubit-level proxy fidelities according to Eq.~\ref{eq:circ_fid}.
We assess the circuit-level fidelity estimation accuracy using three representative circuit BV, GHZ, and random circuits, the results are shown in Fig.~\ref{fig:eval_1_2}. The BV and GHZ circuits are executed on real IBM quantum hardware, while the random circuits are executed on noisy simulators that emulate hardware noise using the corresponding calibration data.

For BV circuits, the ideal output is a single deterministic computational basis state.
Therefore, the ground-truth circuit fidelity is given by the probability of measuring the correct output bitstring. As shown in Fig.~\ref{fig:eval_1_2}(a), the proposed proxy fidelity closely matches the ground-truth fidelity, achieving an AAD of 0.058 and $R^2$ of 0.955.
For GHZ circuits, which generate maximally entangled states $(|0...0\rangle + |1...1\rangle)/\sqrt{2}$, we use the success probability, the sum of the probabilities of observing $|0...0\rangle$ and $|1...1\rangle$, as the reference fidelity. The estimated proxy fidelity closely tracks this success probability, achieving an AAD of 0.069 and an $R^2$ of 0.919 (Fig.~\ref{fig:eval_1_2} (b)).
For random circuits, the ground-truth fidelity is computed directly from the density matrices obtained through noisy simulations. The proxy fidelity achieves the highest estimation accuracy, with an AAD of 0.031 and an $R^2$ of 0.988 (Fig.~\ref{fig:eval_1_2} (c)).

The particularly high accuracy observed for random circuits can be attributed to the fact that both the noisy simulator and the proposed estimator rely on the same hardware calibration data.
This consistency highlights that the reliability model in Eq.~\ref{eq:qubit_fid}, which explicitly tracks the accumulation of individual noise channels, accurately captures noise propagation throughout circuit execution.
The slightly higher estimation error observed for BV circuits is likely due to temporal noise drift in real hardware and the limited precision of calibration data.
Such discrepancies may also partially stem from crosstalk effects, which are not modeled in the current framework.
In contrast, the lower apparent accuracy for GHZ circuits primarily arises from the limitation of the success probability metric itself, which cannot fully reflect coherence loss in highly entangled states.

Overall, these results demonstrate that the proposed proxy fidelity provides a consistent and physically grounded approximation of circuit reliability across different circuit types and execution environments.

\subsection{Ranking Consistency of Reliability Metrics and Fidelity Evaluators}

\begin{figure*}[t]
\centering
\includegraphics[width=1\linewidth]{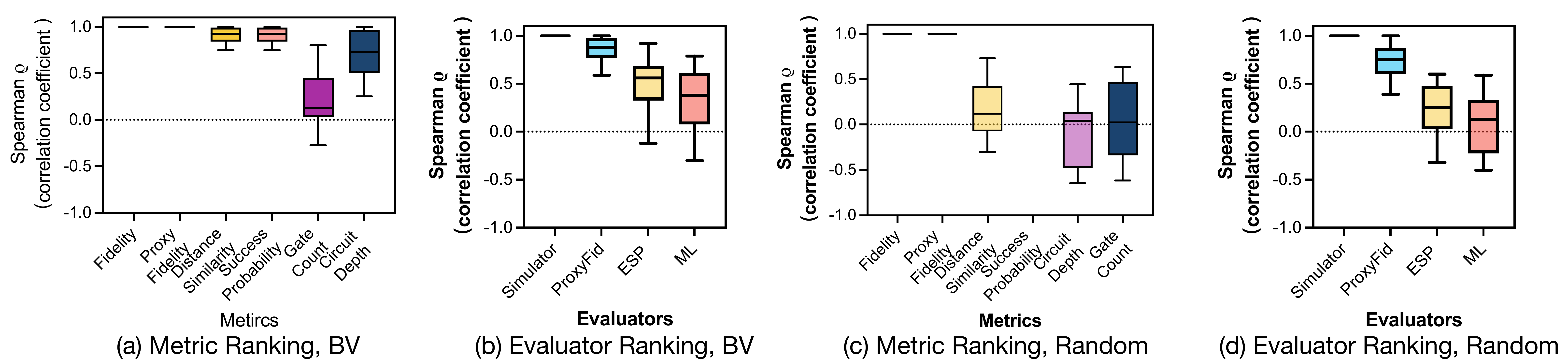}
\caption{Ranking consistency of reliability metrics and fidelity evaluators on BV and random circuits.
Each logical circuit is implemented in 10 different physical qubit layouts. The reliability of each implementation is evaluated using both direct metrics and fidelity evaluators. The consistency of their rankings with the ground truth (actual fidelity ranking) is quantified using Spearman’s correlation coefficient ($\rho$).}
\label{fig:eval_2}
\end{figure*}

We evaluate how well different reliability metrics preserve the ground-truth ordering of circuit implementations, using state fidelity as the reference.
In addition, we assess the ranking accuracy of several fidelity evaluators.
For each evaluator, circuit fidelities are first estimated, circuit implementations are then ranked according to the estimated values, and the resulting rankings are compared with the ground-truth fidelity-based ranking using Spearman’s rank correlation coefficient $\rho$. We consider both BV circuits and random circuits.
For each logical circuit, we generate 10 distinct implementations by varying the logical-to-physical qubit mapping, which leads to different noise accumulation patterns.

The evaluated reliability metrics include \textbf{state fidelity} (ground truth), \textbf{proxy fidelity}, \textbf{distribution similarity} measured as one minus the Hellinger distance, \textbf{success probability} defined as the probability of observing the correct basis state(s), as well as structural metrics including \textbf{circuit depth} and \textbf{gate count}.
In addition, we evaluate several fidelity estimators, including a full noisy-circuit \textbf{Simulator}, the proposed \textbf{ProxyFid} evaluator, \textbf{estimated success probability (ESP)} computed by multiplying per-operation success probabilities, and a trained \textbf{ML}-based evaluator~\cite{liu2020reliability}.

Fig.~\ref{fig:eval_2} (a) shows the ranking performance on BV circuits. 
Proxy fidelity produces the same ranking as state fidelity, since for pure ideal states the state fidelity reduces to the trace inner product between the ideal and noisy states.
Distribution similarity and success probability also achieve high correlation with the ground truth, as the ideal output of BV circuits is a single computational basis state.
In contrast, gate count performs substantially worse than circuit depth, indicating that even coarse structural information provides more reliable guidance than raw operation counts.
These results highlight the importance of circuit structure in reliability evaluation.

Fig.~\ref{fig:eval_2}(c) shows the ranking performance of metrics on random circuits, where success probability is no longer applicable.
Overall, most metrics exhibit degraded performance compared with the BV case, as they fail to capture coherence and interference effects in general quantum states.
In contrast, proxy fidelity continues to closely track the fidelity-based ranking, demonstrating its robustness across circuit types.

Figs.~\ref{fig:eval_2}(b) and (d) report the ranking accuracy of different fidelity evaluators.
The proposed ProxyFid evaluator achieves ranking performance comparable to full noisy simulation, while ESP and the ML-based predictor perform significantly worse, particularly on random circuits.
This gap arises because these methods do not explicitly propagate channel-specific noise effects through the circuit structure.

Across all evaluated circuits, proxy fidelity consistently provides the most fidelity-aligned rankings, and its corresponding evaluator offers an effective and computationally efficient alternative to simulation-based reliability estimation.

\subsection{State-Independence Validation on VQC}

\begin{figure*}[t]
\centering
\includegraphics[width=1\linewidth]{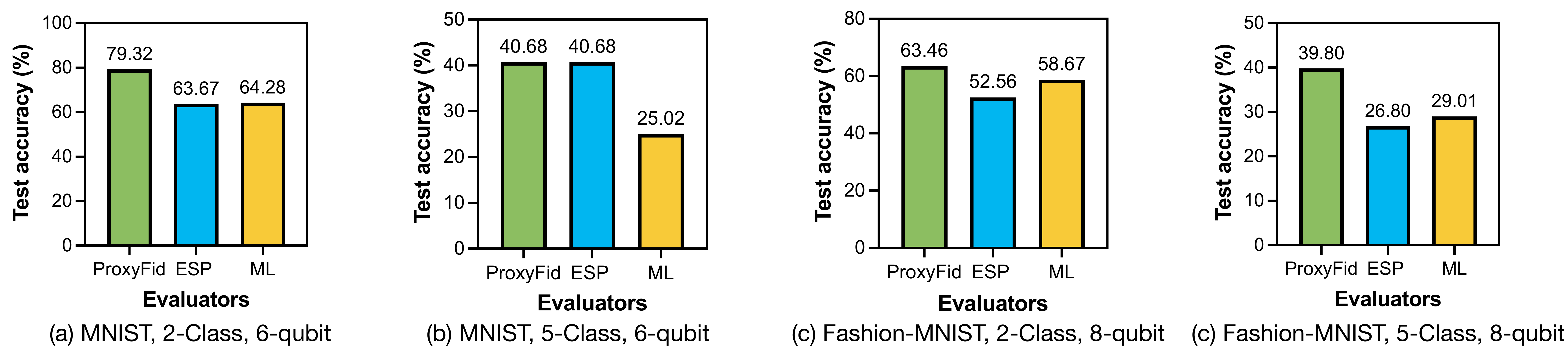}
\caption{Test accuracy under different VQC implementation choices.
For each task, the implementation is selected \emph{prior to training} using different reliability evaluators. ProxyFid consistently selects the implementation that achieves the highest test accuracy across all tasks, outperforming both ESP and ML-based evaluators.
}
\label{fig:eval_3}
\end{figure*}

VQCs are widely used as quantum neural networks. 
Although the gate layout of a VQC is fixed, its quantum state evolution depends on continuously tunable parameters.
This property makes VQCs a natural testbed for evaluating the state-independence of reliability evaluators.

To this end, we generate 10 different implementations of the same VQC by varying the logical-to-physical qubit mapping.
For each implementation, we compute its proxy fidelity \emph{prior to training} and select the top-1 (most reliable) implementation.
For comparison, we perform the same pre-training selection using two alternative evaluators: ESP and an ML-based predictor. We then train the selected implementations on the same dataset using identical model architectures, initial parameters, and hyperparameters.
Under this controlled setting, any performance differences among the trained models are solely attributable to differences in circuit reliability.

We consider both binary (2-class) and multi-class (5-class) classification tasks on the MNIST~\cite{deng2012mnist} and Fashion-MNIST~\cite{xiao2017fashion} datasets. The VQCs use 6 qubits for MNIST and 8 qubits for Fashion-MNIST.
Fig.~\ref{fig:eval_3} reports the test accuracies of the implementations selected by each evaluator.
Across all tasks, proxy fidelity consistently selects the implementation that achieves the highest test accuracy.
Although ESP matches proxy fidelity’s choice in one task, it fails to do so consistently.
The ML-based evaluator performs the worst overall, frequently selecting suboptimal implementations.
These results demonstrate that the proposed proxy fidelity provides an effective and robust \emph{state-agnostic} reliability evaluation for noisy quantum circuits.

\subsection{Case Study with Qubit-level Analysis}

\begin{figure*}[t]
\centering
\includegraphics[width=1\linewidth]{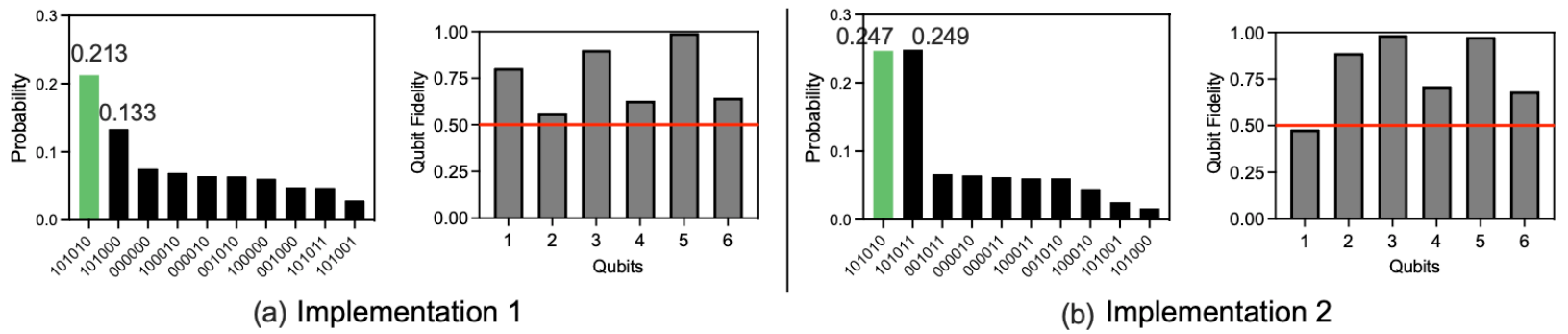}
\caption{Effects of individual-qubit reliability on circuit reliability.
Two implementations of the same BV circuit with the ideal output state $|101010\rangle$ are shown. Implementation 1 (a) is considered less reliable than Implementation 2 (b) according to fidelity and success probability. However, the correct output state can be identified in (a) but not in (b). This conflict arises from the low reliability of qubit 1 in Implementation 2, which causes the circuit to generate an incorrect state with a higher probability than the correct one.
}
\label{fig:fd5}
\end{figure*}

In general, noise-aware circuit implementation evaluates the reliability of candidate layouts at the circuit level. However, our framework not only estimates circuit-level reliability but also provides qubit-level reliability analysis, offering deeper insight into how an individual qubit affects circuit reliability.

Fig.~\ref{fig:fd5} presents two implementations of the same BV circuit with the ideal output state $|101010\rangle$. The state-fidelity metric indicates that Implementation 2 is more reliable than Implementation 1, which is also reflected by its higher probability of producing the correct state in the output distribution. However, the observed output distributions reveal a counterintuitive behavior. In Implementation 1, the correct state can be identified as the one with the highest probability. In contrast, in Implementation 2, an incorrect state appears with a higher probability, making the correct state indistinguishable from measurement results alone.

This conflict arises from the imbalance in qubit reliability. The bar charts on the right show the fidelity of each qubit. In Implementation 1, all qubits have proxy fidelities above 0.5, which is the lowest reliability range before measurement, indicating that each qubit tends to produce its correct state. However, in Implementation 2, qubit 1 exhibits a proxy fidelity below 0.5, suggesting that it is more likely to produce an incorrect value. Consequently, the output distribution of Implementation 2 shows the highest probability for the incorrect state obtained by flipping the first qubit (the most right bit) of the correct bitstring.

This case study underscores the importance of qubit-level reliability analysis for accurately assessing the reliability of quantum systems. Since both local and global reliability contribute to execution outcomes, future work will explore integrating these two levels of analysis into a unified reliability-guided circuit implementation framework.

\section{Discussion} \label{sec:discussion}
The proposed framework aims to provide a state-independent, efficient, and interpretable approach for estimating the reliability of noisy quantum circuits.
Although certain noise processes are inherently state-dependent, our method intentionally adopts a state-agnostic formulation to model how reliability degrades throughout circuit execution.
This abstraction may introduce approximation error relative to full state fidelity.
Nevertheless, by explicitly tracking reliability evolution along individual noise channels at a fine-grained level, the proposed framework achieves substantially higher accuracy and interpretability than existing reliability evaluators while remaining scalable and execution-free.

Our approach relies on hardware calibration data provided by real quantum devices.
However, the current model does not explicitly account for crosstalk errors, as such information is typically absent from standard calibration reports.
Since crosstalk can be approximately modeled as an additional depolarizing channel \cite{gambetta2012characterization}, it can be seamlessly incorporated into our framework once reliable parameterization becomes available.
To further improve estimation accuracy, future work may explore hybrid analytical–learning approaches that integrate partial state information to capture state-dependent noise effects without incurring the full cost of state simulation.

Beyond reliability estimation, efficient, accurate, and interpretable noise modeling has broader implications for noise-aware circuit compilation, reliability-driven optimization, and the design of error mitigation strategies.
By providing transparent insight into how noise accumulates and propagates through a quantum circuit, the proposed framework offers a foundation for building more reliable and scalable quantum computing systems.

\section{Related Work} \label{sec:related}

Reliability evaluation is a fundamental task in noisy quantum computing, as it guides the implementation of quantum circuits on imperfect hardware and directly impacts execution correctness.

Gate-count and depth-based metrics provide the most straightforward way to estimate circuit reliability. The underlying assumption is that fewer gates or shorter circuit depth reduce exposure to noise, thereby improving reliability. Consequently, many compilation techniques aim to minimize gate counts or circuit depth during mapping and optimization \cite{li2019tackling, zhang2020depth, paler2023machine, wu2020qgo, jin2023tetris}. However, these metrics ignore the heterogeneous error rates of quantum operations and fail to accurately reflect the true reliability of circuits executed on real devices.

The Estimated Success Probability (ESP), proposed by Nishio et al. \cite{nishio2020extracting}, has become a widely used reliability metric that incorporates hardware-specific error characteristics. ESP has been integrated into various noise-aware compilation and mapping frameworks \cite{tannu2019ensemble, das2021jigsaw}. Nevertheless, our experimental results show that ESP often fails to accurately rank the reliability of different circuit implementations because it does not consider the circuit structure or the sequential propagation of noise.

Noise-model-based simulation provides a more comprehensive approach for reliability estimation. IBM Qiskit Aer \cite{qiskitaer}, for example, constructs device-specific noise models using calibration data and simulates quantum circuits under realistic noise. Similarly, Dahlhauser et al. \cite{dahlhauser2021modeling} develop application-specific noise models to improve estimation accuracy. While these simulators can accurately reproduce noise effects, both model construction and noisy-state simulation incur significant computational cost. Moreover, the classical simulation of large-scale quantum systems is limited by exponential memory and time complexity, making this approach unscalable.

Recently, machine-learning-based approaches have been explored for predicting circuit reliability \cite{saravanan2021test, liu2020reliability, wang2022quest, zlokapa2020deep, vadali2024quantum}. Liu et al. \cite{liu2020reliability} proposed a neural network model that estimates circuit reliability using features such as circuit depth, width, and gate count. Wang et al. \cite{wang2022quest} introduced a graph transformer model to predict circuit reliability, while Zlokapa et al. \cite{zlokapa2020deep} trained a convolutional neural network to learn device-specific noise models. Although ML-based methods can incorporate hardware and structural information, they typically operate as black-box predictors with limited interpretability. Furthermore, training large-scale models demands substantial computational resources and data collection, limiting their practicality for dynamic and rapidly changing hardware environments.

In contrast, our work proposes a fine-grained, interpretable, and execution-free reliability evaluation framework that analytically models noise accumulation using only calibration data, which are frequently updated to accurately reflect the noise levels of quantum devices. This approach provides fidelity-level accuracy without circuit execution or costly state reconstruction.

\section{Conclusion} \label{sec:conc}
We presented a fine-grained and efficient framework for evaluating the reliability of noisy quantum circuits.
By introducing the Noise Proxy Circuit (NPC) and the Proxy Fidelity metric, the proposed approach estimates circuit reliability without quantum execution or state tomography, explicitly modeling noise accumulation from depolarizing, thermal relaxation, and SPAM channels.
The framework enables state-independent, scalable, and interpretable reliability evaluation using only hardware calibration data. Extensive experiments on BV, GHZ, and random circuits demonstrate that proxy fidelity strongly correlates with true state fidelity, achieving an AAD between 0.031 and 0.069 while substantially reducing computational cost.
These results suggest that the proposed framework provides a practical foundation for reliability-aware quantum circuit implementation on noisy quantum hardware.

% We presented a fine-grained and efficient framework for evaluating the reliability of noisy quantum circuits. The proposed Noise Proxy Circuit (NPC) and Proxy Fidelity metric estimate circuit reliability without execution or tomography by modeling noise accumulation through depolarizing, thermal relaxation, and readout channels. The method achieves state-independent, scalable, and interpretable reliability estimation using only hardware calibration data. Experiments on BV, GHZ, and random circuits show strong correlation with true fidelity, achieving an AAD of 0.031–0.069 while greatly reducing computational cost. 
% This framework provides a practical basis for noise-aware circuit compilation and reliability-driven optimization on NISQ devices.

% \begin{acks}
% \end{acks}

\bibliographystyle{ACM-Reference-Format}

% \bibliography{Article/8_reference.bib}

\appendix
% \section{Appendix}
\appendix

\section{Reproducing Fig. 5} \label{app:fig_5}
This appendix details how the numerical values in Fig.~\ref{fig:method_2} are computed.
Figure~\ref{fig:method_2} illustrates the relationship between qubit proxy fidelity and entanglement degree (measured by negativity) under (a) depolarizing noise and (b) thermal relaxation noise.

\paragraph{Circuit configuration.}
We consider a two-qubit circuit that prepares an entangled state by applying a single-qubit rotation $R_y(\theta)$ on $q_1$, followed by a CNOT gate with $q_1$ as the control qubit and $q_2$ as the target qubit.
By varying $\theta \in \{\pi/8,\ \pi/4,\ \pi/2\}$, the resulting two-qubit state exhibits different degrees of entanglement between $q_1$ and $q_2$.

\paragraph{Entanglement metric.}
We use \textit{negativity} as the entanglement metric~\cite{vidal2002computable}, as it is effective for quantifying entanglement in both pure and mixed quantum states.
Negativity is defined as
\begin{equation}
\mathcal{N}(\rho) \;=\; \frac{\|\rho^{T_i}\|_1 - 1}{2},
\end{equation}
where $\rho^{T_i}$ denotes the partial transpose of the density matrix $\rho$ with respect to the $i$-th qubit, and $\|\cdot\|_1$ denotes the trace norm.
For a two-qubit system, the negativity ranges from $0$ (no entanglement) to $0.5$ (maximal entanglement).
For the ideal states generated by the circuit, the corresponding negativity values are $0.19$, $0.35$, and $0.5$ for $\theta \in \{\pi/8,\ \pi/4,\ \pi/2\}$, respectively.

\paragraph{Depolarizing noise (Fig.~\ref{fig:method_2}(a)).}
To evaluate the effect of depolarizing noise, we apply a depolarizing channel $\mathcal{D}(p)$ to qubit $q_1$, where the depolarizing probability $p$ is swept over the range $[0,1]$ with a step size of $0.01$.
The qubit proxy fidelity of $q_1$ before applying the depolarizing channel is initialized to $1$.
For each value of $p$, we compute (i) the resulting entanglement degree using negativity and (ii) the qubit proxy fidelity using Eq.~\ref{eq:proxyf_dep_iter}.
Each pair of values is plotted as a data point in Fig.~\ref{fig:method_2}(a) (right).

\paragraph{Thermal relaxation noise (Fig.~\ref{fig:method_2}(b)).}
To evaluate the effect of thermal relaxation noise, we apply a thermal relaxation channel $\mathcal{T}(t)$ to qubit $q_1$, where the operation duration $t$ is swept over the range $[0,T_2]$ with a step size of $0.01$.
The qubit proxy fidelity of $q_1$ before applying the thermal relaxation channel is likewise initialized to $1$.
For each value of $t$, we compute (i) the resulting entanglement degree using negativity and (ii) the corresponding qubit proxy fidelity using Eq.~\ref{eq:trc}.
Each pair of values is plotted as a data point in Fig.~\ref{fig:method_2}(b) (right). \\

Overall, the results in Fig.~\ref{fig:method_2} show that negativity and the computed qubit proxy fidelity remain positively correlated under both depolarizing and thermal relaxation noise.
This consistent relationship demonstrates that the proposed proxy fidelity effectively reflects noise-induced variations in inter-qubit correlations.

\section{Decompose multi-qubit depolarizing channel} \label{app:2q_dep}
The error of a multi-qubit noisy gate acts as a multi-qubit depolarizing channel. To analyze its effect on individual qubits, an $n$-qubit depolarizing channel can be approximated by $n$ parallel single-qubit depolarizing channels. Each channel independently affects its corresponding qubit with the same depolarizing parameter as the original multi-qubit channel, as illustrated in Fig.~\ref{fig:method_3} (c), bottom.

% Fig.~\ref{fig:fd1_2} illustrates the decomposition of a 2-qubit depolarizing channel.
Specifically, consider a two-qubit state represented by the density matrix \( \rho_{AB} \):
\begin{equation*}
    \rho_{AB} = \begin{bmatrix}
    a & b & c & d \\
    b^* & e & f & g \\
    c^* & f^* & h & i \\
    d^* & g^* & i^* & j
\end{bmatrix}
\end{equation*}
The density matrices of the individual qubits are \( \rho_A = \text{Tr}_B(\rho_{AB}) \) and \( \rho_B = \text{Tr}_A(\rho_{AB}) \), respectively:
\begin{equation*}
\rho_{A} =  \begin{bmatrix}
a+e & c+g \\
c^*+g^* & h+j
\end{bmatrix}
\quad
    \rho_{B} = 
\begin{bmatrix}
a+h & b+i \\
b^*+i^* & e+j
\end{bmatrix}
\end{equation*}

After applying a 2-qubit depolarizing channel with parameter \( p \) to \( \rho_{AB} \), we obtain \( \rho_{AB}' = \mathcal{D}_{\text{depol}}(\rho_{AB}, p) \):
\begin{equation*}
    \rho'_{AB} =  \begin{bmatrix}
    (1-p)a+\frac{p}{4} & (1-p)b & (1-p)c & (1-p)d \\
    (1-p)b^* & (1-p)e+\frac{p}{4} & (1-p)f & (1-p)g \\
    (1-p)c^* & (1-p)f^* & (1-p)h+\frac{p}{4} & (1-p)i \\
    (1-p)d^* & (1-p)g^* & (1-p)i^* & (1-p)j+\frac{p}{4}
\end{bmatrix}
\end{equation*}
The current states of the qubits are \( \rho_A' = \text{Tr}_B(\rho_{AB}') \) and \( \rho_B' = \text{Tr}_A(\rho_{AB}')\):
\begin{equation*}
\rho'_{A} = \begin{bmatrix}
(1-p)(a+e)+\frac{p}{2} & (1-p)(c+g) \\
(1-p)(c^*+g^*) & (1-p)(h+j)+\frac{p}{2}
\end{bmatrix}
\end{equation*}
\begin{equation*}
    \rho'_{B} = 
\begin{bmatrix}
(1-p)(a+h)+\frac{p}{2} & (1-p)(b+i) \\
(1-p)(b^*+i^*) & (1-p)(e+j)+\frac{p}{2}
\end{bmatrix}
\end{equation*}
These states are equivalent to each qubit passing through a single-qubit depolarizing channel with the parameter \( p \), such that \( \rho_A' = \mathcal{D}_{\text{depol}}(\rho_A, p) \) and \( \rho_B' = \mathcal{D}_{\text{depol}}(\rho_B, p) \). 
Therefore, a 2-qubit depolarizing channel can be decomposed into two single-qubit depolarizing channels, as illustrated in Fig.~\ref{fig:method_3} (c), bottom.
% \begin{figure}[h]
% \centering 
% \includegraphics[width=0.55\columnwidth]{fidelityDynamics/figures/FD1_2.pdf}
% \caption{Decomposition of a 2-qubit Depolarizing Channel}
% \label{fig:fd1_2}
% \end{figure}


\end{document}